\documentstyle[prl,multicol,aps,epsfig]{revtex}

\begin{document}

    \title{Neutron Stars\footnote{To be published in {\em Encyclopedia of
Physics 3rd ed.}, R.G.  Lerner and G.L.  Trigg, eds., Wiley-VCH, Berlin.
Research supported in part by NASA grant NAG~5-12030, NSF grants PHY~03-55014
and AST~0098399, and the Fortner Endowed Chair at the University of Illinois.
}}
\author{Gordon Baym and Frederick K. Lamb}
\address{Department of Physics, University of Illinois at
Urbana-Champaign, 1110 West Green Street, Urbana, Illinois 61801}
\date{\today}
\maketitle

\begin{abstract}
\end{abstract}

\begin{multicols}{2}
\narrowtext

    Neutron stars are highly condensed stellar objects produced in supernova
explosions, the end point in the evolution of more massive stars.  Neutron
star masses $M$ are in the range of 1-2 times that of the sun ($M_\odot$),
whereas typical radii are only {10-12}{km}; the matter they contain, primarily
neutrons, is thus the densest found in the universe today.  The average
interior density is greater than that in a large atomic nucleus, $\rho_0 =
{3\times10^{14}}$g/cm$^3$.  (By comparison, white dwarf stars, which are of
similar masses, have radii of at least several thousand kilometers, and
central densities in the range of $10^5-10^9${g/cm$^3$}.)  Support against
gravitational collapse in a neutron star is provided by the quantum-mechanical
Fermi (or zero-point) pressure of the neutrons and other particles in the
interior, in the same way that white dwarfs are supported by electron
zero-point pressure.  (Ordinary stars, on the other hand, are supported by
thermal gas pressure.)

    Neutron stars were first proposed by Baade and Zwicky in 1934 in their
pioneering paper on supernovae, and considerable theoretical work on their
properties, beginning with calculations by Oppenheimer and Volkoff in 1939,
was carried out prior to their actual observation.  It was not until the
discovery in 1967 by Bell and Hewish of radio pulsars -- stars whose radio
emission appears to blink on and off -- and their identification by Gold as
rotating neutron stars, that the existence of neutron stars was established.
Since that time neutron stars have become cosmic laboratories for testing
fundamental physics, including relativistic theories of gravity and the
properties of matter at extreme densities.  Neutron stars also play a crucial
astrophysical role as the objects underlying a wide variety of highly
energetic compact radio, x-ray, and gamma-ray sources.  Radio pulsars are
rotation-powered, are found both in isolation and in binary star systems, and
are observed to emit radiation at all frequencies from radio to optical to
gamma rays.  Neutron stars are also found in luminous compact x-ray binaries
in which they accrete matter from a companion star.  About 1500 neutron stars
have so far been detected in the galaxy as radio pulsars, including about 125
such pulsars with millisecond periods.  More than 200 accretion powered
neutron stars have been detected in x-ray binary systems; about 50 are x-ray
pulsars and a similar number produce intense x-ray bursts powered by
thermonuclear flashes.

    The astrophysical energy source of neutron stars, some 10~times as
powerful as thermonuclear burning, is their immense gravitation.  The
gravitational acceleration ($g=GM/R^2$, where $G$ is Newton's gravitational
constant) at the surface is about $10^{11}$ times that on Earth; gravitational
tidal forces would make it impossible for any normal object larger than about
{10}{cm} to reach the surface of a neutron star without being torn apart.  The
gravitational binding energy $GmM/R$ of a particle of mass $m$ at the surface
is about one-tenth of its rest energy $mc^2$.  (Nuclear binding energies are,
in comparison, at most 0.9\% of the rest energy of matter.)  The energy
emitted by neutron stars in x-ray binary systems comes primarily from the
gravitational energy released by matter accreted onto the neutron star from
its companion star; nuclear energy contributes a few percent.  The energy
source of rotating pulsars -- the kinetic energy of rotation of the neutron
star -- also comes from the release of gravitational binding energy, for as
the stellar core in a supernova collapses under gravity to form a neutron
star, conservation of angular momentum requires that its rotation rate and
rotational energy increase.  Other neutron stars gradually release
gravitational energy stored as magnetic or thermal energy.

    Neutron stars characteristically are strongly magnetized, with surface
magnetic fields ranging from {10$^{6}$}{G} to {10$^{15}$}{G}.  The slowing
down of pulsar rotation implies magnetic fields of order {10$^{11}$}{G} to
{10$^{15}$}{G}.  The rapid rotation of such fields is important in generating
relativistic particles and radio emission.  Plasma accreting onto neutron
stars in x-ray binary systems is channeled to the magnetic poles by fields
ranging from $10^8$ to {10$^{13}$}{G}.

    At birth, the interior temperature of a neutron star is about
{10$^{11}$}{K}, and within the first few days it cools by neutrino emission to
less than {10$^{10}$}{K}.  Throughout most of its early life, the interior
temperature is in the range $10^8-{10^9}${K} and the surface temperature is
one-tenth or less of the interior temperature.  The hot surface typically
radiates x-rays.  X-ray satellites such as Chandra and XMM Newton have
detected x-ray emission from cooling stars but the emission process and the
size and structure of the emitting regions are not as yet sufficiently
understood to infer accurately neutron star surface temperatures and radii
from these measurements.

    The matter inside neutron stars has a relatively low temperature compared
with its characteristic microscopic energies of excitation, typically of the
order of MeV (million electron volts).  Furthermore, nuclear processes in the
early moments of a neutron star take place sufficiently rapidly compared with
the cooling of the star that the matter essentially comes -- via strong and
electromagnetic interactions, as well as weak interactions (which transform
protons into neutrons and vice versa) -- into its lowest possible energy
state.

    The cross section of a neutron star interior is shown in Fig.~1.  The
density of matter increases with increasing depth in the star.  Beneath an
atmosphere, compressed by gravity to less than {1}{cm} height, is a crust,
typically {$\sim$1}{km} thick, consisting, except in the molten outer tens of
meters, of a lattice of bare nuclei immersed, as in a normal metal, in a sea
of degenerate electrons.  The matter in the outer part of the crust is
expected to be primarily $^{56}$Fe, the end point of thermonuclear burning
processes in stars.

\begin{figure}
\begin{center}
\epsfig{file=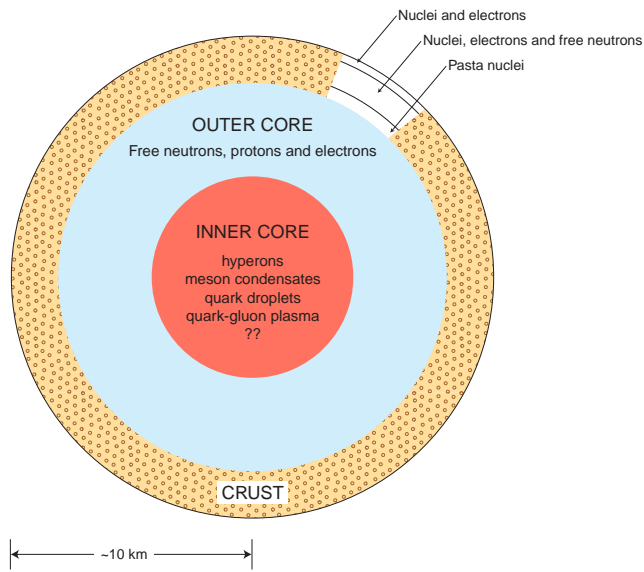,height=7.5cm}
\end{center}
\caption
{Schematic cross section of a neutron star, showing the outer
crust consisting of a lattice of nuclei with free electrons, the inner crust
which also contains a gas of neutrons, the nuclear ``pasta" phases, the
liquid outer core, and the possibilities of higher-mass baryons,
Bose-Einstein condensates of mesons, and possible quark matter in the inner
core. (Figure kindly drawn by Michael Baym.)}
\label{FIG1}
\end{figure}

    With increasing depth, the electron Fermi (or zero-point) energy rises.
Beyond the density {8$\times10^6$}{g/cm$^3$}, it is so high ($>1${MeV}) that
$^{56}$Fe can capture energetic electrons.  In the capture process, as occurs
during the formation of the neutron star in a supernova, protons in nuclei are
converted into neutrons via the weak interaction
$e^-+{\mathrm{p}}\to{\mathrm{n}} + \nu$.  The produced (electron) neutrino
$\nu$ escapes the nascent neutron star, lowering the energy of the system.
(Neutrinos generated in the formation and neutronization of the neutron star
are eventually responsible for ejection of the mantle in the supernova.)  The
matter becomes more neutron rich and rearranges into a sequence, with
increasing density, of increasingly neutron-rich nuclei, reaching nuclei such
as $^{118}$Kr at a mass density
$\rho_{\mathrm{d}}={4.3\times10^{11}}${g/cm$^3$}.  The uncertainties in the
actual nuclei present reflect current uncertainties whether the shell
structure of highly neutron rich nuclei is determined by the same {\em magic
numbers} (e.g., 82 neutrons) that determine the shell structure of normal
nuclei.  The nuclei present deep in the crust, although unstable in the
laboratory, cannot undergo beta decay via the inverse reaction
${\mathrm{n}}\to{\mathrm{p}} + e^- +\bar{\nu}$ (where $\bar{\nu}$ is an
anti-(electron) neutrino) because the electron would, by energy conservation,
have to go into an already occupied state, a process forbidden because
electrons obey the Pauli exclusion principle.  Beyond the density
$\rho_{\mathrm{d}}$, called the ``neutron drip'' point, the matter becomes so
neutron rich that not all the neutrons can be accommodated in the nuclei, and
the matter, still solid, becomes permeated by a sea of free neutrons in
addition to the sea of electrons.  Then at about a density $\sim \rho_0/3$,
spherical nuclei become unstable, as in fission, and the matter proceeds
through a sequence of rather unusual structures, termed ``pasta nuclei," with
the nuclei first becoming rod-like and then laminar, with pure neutrons
filling the space between.  The pure-neutron plates become thinner; eventually
the neutrons form rods, and then spheres, with the between regions containing
proton-rich matter.  Remarkably, over {\it half} the matter in the crust is in
the form of these non-spherical configurations.  Finally, at a density of
about $\rho_0$, the matter dissolves into a uniform liquid composed primarily
of neutrons with a few percent protons and electrons and a smaller fraction of
muons.  The neutrons are most likely superfluid and the protons
superconducting; the electrons, however, are normal.

    The states of matter at high pressures deep in the interior are less well
understood.  With increasing density heavier baryons can live stably in the
star.  Several interesting phenomena are possible (see Fig.~1).  One is that
pi mesons are spontaneously produced and form a superfluid ``Bose-Einstein
condensed'' state; such condensation would greatly enhance the cooling of
neutron stars by neutrino emission.  The matter may similarly undergo an
analogous ``kaon condensation.''  At ultrahigh densities, where the nucleons
are strongly overlapping, matter is expected to dissolve into ``quark
matter,'' in which the quarks that make up the baryons become free to run
throughout.  Quark matter most likely first appears as droplets in a sea of
nuclear matter at densities of order several times $\rho_0$; a core of bulk
quark matter, which would be present at higher densities, would also enhance
neutron star cooling, but whether the transition to bulk quark matter is
actually reached in neutron stars remains uncertain.

    The structure of neutron stars, including their radii as a function of
mass, and the range of masses for which they are stable, is determined by the
equation of state of the matter they contain.  A knowledge of the maximum mass
$M_{\mathrm{max}}$ that a neutron star can have is important in distinguishing
possible black holes from neutron stars by observations of their masses.  The
uncertainty in the present theoretical limit,
$M_{\mathrm{max}}\sim2.5M_\odot$, calculated on the basis of physically
plausible equations of state, reflects our uncertainty of the properties of
matter at densities much greater than $\rho_0$.

    The ever-growing body of observational information on neutron stars
provides increasingly stringent constraints on the structure of neutron stars
and the properties of neutron star matter.  The neutron stars detected in
relativistic double neutron star binary systems have masses between 1.33 and
1.45 $M_\odot$, whereas those of some neutron stars in other compact binary
systems are at least 1.6 to 1.7 $M_\odot$.  The masses of neutron stars in
x-ray binary systems can be estimated by combining optical and x-ray
observations of these systems; the estimated masses range from about one to
$\sim1.9M_\odot$.  Measurements of the frequencies of the kilohertz
quasi-periodic x-ray oscillations of more than 20 accreting neutron stars
potentially constrain simultaneously the masses and radii of these stars.
Information on the moments of inertia of neutron stars can be extracted from
measurements of the behavior of their spin periods and luminosities over time
and from precision measurements of general relativistic spin-orbit coupling in
double neutron star binary systems.  In addition, sudden speedups and smaller
fluctuations in pulsar repetition frequencies provide clues to the internal
structure of neutron stars, and support the idea that the interior is
superfluid.

See also: {Fermi--Dirac Statistics; Pulsars; Stellar Energy Sources and
Evolution.}

\begin{description}

    \item G. Baym and C.J. Pethick, \textit{Annu.\ Rev.\ Nucl.\ Sci.}
\textbf{25}, 27 (1975); \textit{Annu.\ Rev.\ Astron.\ Astrophys.} \textbf{17},
415 (1979); H. Heiselberg and V.R.  Pandharipande, \textit{Annu.\ Rev.\ Nucl.\
Part. \ Sci.} \textbf{50}, 481 (2000).  (On properties of matter in neutron
stars; extensive references.)  (A)

    \item C.L. Fryer in \textit {The neutron star--black hole connection},
Proc.\ NATO Advanced Study Institute, NATO science series C, Vol. 567 (C.
Kouveliotou, J. Ventura, and E.P.J. van den Heuvel eds.), Kluwer, Dordrecht,
2001.  (On formation of neutron stars in supernovae.)

    \item F.K. Lamb and W. Yu, in {Binary Radio Pulsars}, ASP Conference
Series no. 328 (F.  Rasio and I. Stairs, eds.), ASP, San Francisco, CA, 2005.
(Summarizes current knowledge of neutron star magnetic fields and spin rates,
in a useful volume on the properties of x-ray and radio pulsars and their use
in testing relativistic theories of gravity.)

    \item W.H.G.  Lewin and M. van der Klis (eds.), \textit {Compact Stellar
X-ray Sources}.  Cambridge Univ.  Press, 2005.  (Overview of the properties of
neutron stars in x-ray binary systems.)

    \item A.G. Lyne and F. Graham-Smith, \textit {Pulsar Astronomy}, 3rd. ed.
Cambridge Univ.  Press., 2005.  (Describes supernovae, x-ray binary systems,
isolated and binary radio and x-ray pulsars, and x-ray bursters.)

    \item D.J. Nice, E.M. Splaver, \& I.H. Stairs, in \textit {Young
Neutron Stars and Their Environments}, IAU Symposium, Vol. 218 (F.  Camilo and
B.M. Gaensler, eds.), 2004; S. M. Ransom, et al., Science, 307, 892, 2005.
(Evidence for massive neutron stars.)

    \item S.L. Shapiro and S.A. Teukolsky, \textit{Black Holes, White Dwarfs
and Neutron Stars}.  Wiley, New York, 1983.  (Comprehensive text on physics of
compact stars and black holes.)  (I)

\end{description}

\end{multicols}
\end{document}